\documentclass[conference]{IEEEtran}
\IEEEoverridecommandlockouts
\usepackage{hyperref}
\hypersetup{colorlinks,allcolors=black}
\usepackage{cite}
\usepackage{amsmath,amssymb,amsfonts}
\usepackage{algorithmic}
\usepackage{graphicx}
\usepackage{textcomp}
\usepackage{xcolor}
\usepackage{float}
\usepackage{diagbox}
\usepackage{soul}
  
\def\BibTeX{{\rm B\kern-.05em{\sc i\kern-.025em b}\kern-.08em
    T\kern-.1667em\lower.7ex\hbox{E}\kern-.125emX}}
\begin{document}

\title{Comparison of Deep Learning and Traditional Machine Learning Techniques for Classification of Pap Smear Images\\
}

\author{\IEEEauthorblockN{1\textsuperscript{st} Abdurrahim Yilmaz}
\IEEEauthorblockA{\textit{Yildiz Technical University} \\
\textit{Mechatronics Engineering}\\
Istanbul, Turkey \\
a.rahim.yilmaz@gmail.com}
\and
\IEEEauthorblockN{2\textsuperscript{nd} Ali Anil Demircali}
\IEEEauthorblockA{\textit{Yildiz Tehnical University} \\
\textit{Mechatronics Engineering}\\
Istanbul, Turkey \\
aanil@yildiz.edu.tr}
\and
\IEEEauthorblockN{3\textsuperscript{rd} Sena Kocaman}
\IEEEauthorblockA{\textit{Robert College} \\
\\
Istanbul, Turkey \\
senakocaman@gmail.com}
\and
\IEEEauthorblockN{4\textsuperscript{th} Huseyin Uvet}
\IEEEauthorblockA{\textit{Yildiz Tehnical University} \\
\textit{Mechatronics Engineering}\\
Istanbul, Turkey \\
huvet@yildiz.edu.tr}
}

\maketitle

\begin{abstract}
A comprehensive study on machine and deep learning techniques for classification of normal and abnormal cervical cells by using pap smear images from Herlev dataset results are presented. This dataset includes 917 images and 7 different classes. All techniques used in this study are modeled by using Google Colab platform with scikit-learn and Keras library inside TensorFlow. In the first study, traditional machine learning methods such as logistic regression, k-Nearest Neighbors (kNN), Support Vector Machine (SVM), Decision Tree, Random Forest and eXtreme Gradient Boosting (XGBoost) are used and compared with each other to find binary classification as normal and abnormal cervical cells. Better results are observed by XGBoost and kNN classifiers among the others with an accuracy of 85\%. In the second study, a deep learning model based on Convolutional Neural Network(CNN) is used for the same dataset. Accordingly, accuracies of 99\% and 93\% are obtained for the training and the test dataset, respectively. CNN model extracts its features from raw data without any label or a feature dataset. In this model, it takes 50 epochs to have these accuracies within 20 minutes of computational time.
\end{abstract} 


\section{Introduction}
Cervical cancer is one of the most influential cancer type for women in worldwide according to Global Cancer Observatory. It is reported that 570,000 new cases and 311,000 deaths are occurred in 2018 due to this cancer \cite{bray2018global}. It is important to find a related diagnosis and find a cure; therefore, machine and deep learning approaches can be used to detect various disease types such as cervical cancer, onychomycosis, breast cancer and etc. by using robust models \cite{bejnordi2017diagnostic, han2018deep}. To do this, a dataset must be generated for automated illness classification. For automated classification of cervical cancer, Herlev dataset is created from pap smear images and each image of this dataset has been labeled initially. Thus, it can be used for both machine and deep learning techniques since machine learning techniques require labeled images for training. On the other hand, in this dataset, seven cervical cell types are available such as superficial squamous epithelial, intermediate squamous epithelial, columnar epithelial, mild dysplasia, moderate dysplasia, severe dysplasia, and carcinoma-in-situ. These types must be converted from seven types to binary classes as abnormal and normal cells for the classification of seven classes. In this dataset, abnormal cervical cells are mild dysplasia, moderate dysplasia, severe dysplasia, and carcinoma-in-situ. In the meantime, normal cervical cells are superficial squamous epithelial, intermediate squamous epithelial, columnar epithelial \cite{jantzen2005pap}.

Jantzen et al. \cite{jantzen2005pap} studied the classification of Herlev dataset with linear least-squares classifier with an accuracy of 94\%. However, this accuracy stands lower for current methods that have better metrics by using pretrained CNN especially deep learning algorithms \cite{zhang2017deeppap}.
Marinakis et al. \cite{marinakis2009pap} classified cervical cells by using genetic algorithm for segmentation and meta-heuristic algorithm and nearest neighbor based classification with 20 numerical features. Indicated results with an accuracy of 89\% showed that this accuracy is not adequate compared with the current machine learning techniques such as the fuzzy C-means method \cite{chankong2014automatic}. 
Chankong et al. \cite{chankong2014automatic} developed a novel method to segmentate for pap smear images with an accuracy of 99\%. This accuracy is achieved by using patch-based fuzzy C-means (FCM) and FCM clustering method for segmentation and FCM algorithm for classification with features that are obtained from cytoplasm, nucleus and background. Although this is the highest accuracy for binary classification of pap smear images by using the same dataset, the data preprocessing and the segmentation process take a long time and are challenging processes.
Ashok et al. \cite{ashok2016comparison} used multi-thresholding method for segmentation and SVM classifier with 99\% accuracy for a private dataset with 14 texture and 30 shape features. The features are obtained for a different dataset that includes 150 images. Though, the accuracy level is higher, no information is available regarding to complicated and larger dataset.
Zhang et al. \cite{zhang2017deeppap} proposed a deep convolutional neural networks for classification of pap smear images with (256 $\times$ 256 pixels) image size. The proposed method uses transfer learning by using ImageNet dataset to set the weights of deep neural network efficiently. To the best of our knowledge, this model has 98\% accuracy with the highest accuracy for the studies related with pap smear classification. In deep learning, feature extraction is an automated process unlike machine learning. This property reduces the model developing process. However, average training time progresses slowly, which takes 4 hours for 30 epochs and 256 $\times$ 256 pixels image size in this study for each training effort.

The significance of this paper is to show how the models are tested and improved through optimization and reduced computing time using Google Colab platform. Preprocessing and classification stages for machine learning takes a small amount of time, which is around 20 seconds for 1000 hyper-parameter combination to have better accuracy. Deep learning model also takes 20 minutes with 93\% accuracy and 64 $\times$ 64 image size without any transfer learning or pretrained model. In short, Google Colab platform is very efficient for developing machine learning models which can be developed that have high accuracy for a short time as other platform that requires any expensive hardware.
\section{Methodology}
In this section, algorithms and important hyperparameters selection criterias are explained in detail. 7 different machine learning algorithms such as logistic regression, kNN, SVM, Naive Bayes, Random Forest, XGBoost and a deep neural network (CNN) are presented. Hyperparameters and their effects are different for each algorithm and dataset in the model. The most used open access dataset that is Herlev dataset is selected to compare between ours and others' methodology and hyperparameters. The dataset has 20 morphological features with its labels that are important for machine learning algorithms. Thereby, the performance of models is compared for same condition by using same features. Results of all developed models are measured by using scikit-learn library. The obtained hyperparameters and results of the models in which conditions are mentioned. Especially, computational time of training of the CNN model is reduced to 22 seconds for each epochs. The following subsections include detailed information about models and their parameters.

\subsection{Machine Learning}
Traditional machine learning algorithms are used in many areas like regression, classification. The machine learning algorithms use features of data to develop a model. At the same time, feature selection is a crucial part for machine learning, as the features must be distinctive, and changes between the measurements for each specific cell classes must be non-negligible and measurable. The distinctive features increase the performance of the model for classification with the highest metrics. 
Thereby, these features have significant properties of cervical cells, which are nucleus and cytoplasm area with their ratio, brightness, shortest and longest diameter, elongation, perimeter, position, maxima, minima and roundness. Logistic regression, k-Nearest Neighbours k-NN, SVM with different kernels, Naive Bayes, Decision Trees, Random Forest and XGBoost classifiers are used for classifications of normal and abnormal cervical cells by using features of the dataset.
\subsubsection{Logistic Regression}
Logistic regression uses the logistic function to classify data especially in statistical analysis since it helps to understand the relation between the probability of a variable and a feature of a dataset \cite{hosmer2013applied}. Categorical classification for this is convenient. In this study, logistic regression is used to observe the performance of the classification of pap-smear images as abnormal and normal cells.

\subsubsection{k-Nearest Neighbors Classifier}
kNN is a supervised and lazy machine learning algorithm which is capable of memorizing the dataset. In this way, it is a robust algorithm for an old dataset \cite{peterson2009k}. However, it requires a larger memory size since, it keeps all states in calculating distances. In this study, the highest accuracy is observed by using the number of neighbors, the distance metric and the power parameter is selected 9, Minkowski, Euclidean distance respectively.

\subsubsection{Support Vector Machine Classifier}
SVMs are a supervised machine learning algorithm based on the statistical learning theory. SVMs are used in many areas like face, optical character, voice recognition with many advantages like high efficiency for a high dimensional dataset that the algorithm splits data with the best line \cite{cortes1995support}. Because of the advantages of kernels, the kernel of SVM is specified as radial basis function (rbf) for the highest accuracy.

\subsubsection{Naive Bayes Classifier}
Naive Bayes classifier is aimed to classify the data according to probability principles, the ground of Bayes Theorem. It calculates all the probabilities for each data and classifies by the highest probability. It is used in many areas like spam filtering, so Gaussian Naive Bayes classifier is considered for our work \cite{rish2001empirical}.

\subsubsection{Decision Tree Classifier}
A decision tree is a method that uses a series of decision rules to split data into smaller clusters. This approach is an old but efficient tree-based algorithm that is the fundamentals of some algorithms like random forest. For this work, the function to measure the quality of a split is based on the entropy criterion \cite{quinlan1986induction} also, to see entropy of the dataset.

\subsubsection{Random Forest Classifier}
Random forests are an advanced tree-based algorithm that choose subsets from both the features and the dataset to avoid overfitting. Its advantage is that random forest can produce a high accuracy without hyperparameter optimization. Also, the probability of over-fitting is lower than decision trees. However, the number of trees in random forest is a very important parameter and it should be selected well \cite{breiman2001random}.

\subsubsection{XGBoost Classifier}
XGBoost is a very popular and a new machine learning algorithm which is very efficient since it wins half of the challenges in machine learning platform like Kaggle \cite{nielsen2016tree}. XGBoost is a decision tree based and a gradient boosted algorithm. It has many systems and algorithmic optimizations like tree pruning, parallelization, and cross-validation \cite{chen2015xgboost}.

\begin{figure*}[t]
    \centering
    \includegraphics[width=0.8\textwidth]{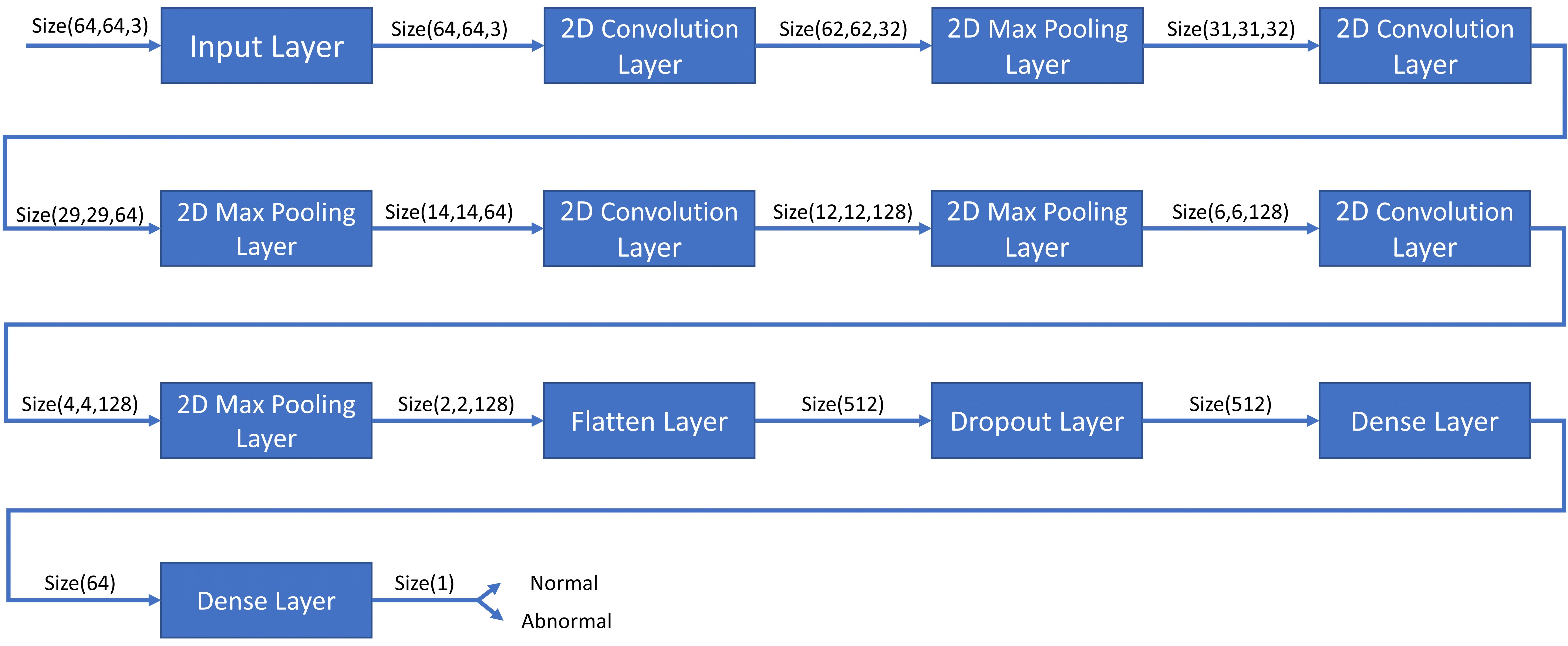}
    \caption{Shows the layer and size of data of convolutional neural network model for pap smear image classification.}
    \label{fig:p1}
\end{figure*}%

\subsection{Deep Learning}
CNN is based on convolution integral where as in the convolution layer, which includes an activation function like Rectified Linear Unit (ReLU) to add nonlinearity to the model \cite{lecun1995convolutional}. In this work, a standard CNN is used for classification and analysis of visual imagery. Image size, train-test split ratio, convolutional layer number, dropout and epoch number are some properties of CNN. The properties of neural network and the dataset that are used in this work are shown in Table \ref{tab:p2}. In addition to this, the block diagram of CNN is shown in Figure \ref{fig:p1}. The model is developed with Keras with 4 convolution layer.

\begin{table}[b!]
\renewcommand{\arraystretch}{1.25}
\centering
\caption{Properties of the CNN}
\begin{tabular}{ccc}
\hline\hline      
\textbf{Property}      & \textbf{Value}      \\ 
\hline  
Image Size           & (64,64,3)             \\
Test Data/All Data   & 0.15                  \\
Convolutional Layer Number   & 4              \\
Dropout                     & 0.4                 \\
2D Pooling Size             & 2x2        \\
Validation Data/Training Data  & 0.15                   \\
Epoch Number         & 50                   \\
\hline\hline
\end{tabular} \label{tab:p2}
\end{table}

\subsection{Metrics}
The performance of models is measured and compared with each other by using metrics. The metrics show accuracy, recall, precision, specificity and $F_1$ score that are calculated by using a confusion matrix. The confusion matrix includes the essential parameter of metrics. The metrics can be calculated by using the essential parameters  according to following criterias:

\begin{itemize}
    \item True Positive(TP) value is an outcome where the model predicted the dataset positive in a correct way.
    
    \item True Negative(TN) value describes how many negative values are predicted correctly.
    
    \item False Positive(FP) value represents that predicted value is positive but it is not. 
    
    \item False Negative(FN) value represents that negative values are predicted wrongly.
\end{itemize}

These are essential metrics and other metrics like accuracy that give more information about models that are obtained by using TP, TN, FP and FN values. 

\begin{itemize}
    \item The accuracy is the rate of true samples among all samples. Higher accuracy means a higher rate of correctly classified data.
    
    \item The recall value is the rate of correctly predicted positive samples among all actual positive samples. Higher recall value means lower misclassified positive data.
    
    \item The precision value is the rate of correctly predicted positive samples among all positive predicted samples. Higher precision means higher correct classified data for true results.
    
    \item The specificity value is the rate of correctly predicted negative results among all actual negative samples. Higher specificity means higher correct classified data for negative results.
    
    \item $F_1$ score is a harmonic mean of recall and precision. It prevents choosing the wrong model unless the dataset is correctly splitted.
\end{itemize}    

\begin{table}[b!]
\renewcommand{\arraystretch}{1.25}
\centering
\caption{Metrics}
\begin{tabular}{ccc} 
\hline\hline
\textbf{Metric}  & \textbf{Formula}              \\ 
\hline \\ [-10pt]
Accuracy       &   $\dfrac{T P+T N}{T P+F P+T N+F N}$  \\ [5pt]
Recall  & $\dfrac{T P}{T P+F N}$ \\ [5pt]
Precision  & $\dfrac{T P}{T P+F P}$ \\ [5pt]
Specificity  & $\dfrac{T N}{T N+F P}$ \\ [5pt]
$F_1$ Score  & $2*\dfrac{\text{Precision} \times \text{Recall}}{\text{Precision}+\text{Recall}}$ \\ [5pt]
\hline\hline
\end{tabular} \label{tab:p3}
\end{table}

\begin{table*}
\renewcommand{\arraystretch}{1.25}
\centering
\caption{Results of Metrics of Machine and Deep Learning Approaches}
\label{tab:p4}
\begin{tabular}{c|ccccccccc} & \multicolumn{9}{c}{ \textbf{Result} }  \\
\hline
\multicolumn{1}{l|}{\diagbox{\textbf{Metrics}}{\textbf{Name}}} &
\begin{tabular}[c]{@{}c@{}} Logistic\\ Regression \end{tabular} & \begin{tabular}[c]{@{}c@{}} k-NN \end{tabular} & \begin{tabular}[c]{@{}c@{}} SVM \end{tabular} & \begin{tabular}[c]{@{}c@{}} Naive Bayes\\ Classifier \end{tabular} & \begin{tabular}[c]{@{}c@{}} Decision \\ Tree \end{tabular} & \begin{tabular}[c]{@{}c@{}} Random\\Forest \end{tabular} & \begin{tabular}[c]{@{}c@{}} XGBoost\\Classifier \end{tabular} & \begin{tabular}[c]{@{}c@{}} CNN for\\Training Data \end{tabular} & \begin{tabular}[c]{@{}c@{}} CNN for\\Test Data \end{tabular}  \\
\hline
Accuracy (\%)    & 83 & 85 & 83 & 80 & 80 & 83 & 85 & 99 & 93 \\
Recall (\%)     & 83 & 87 & 85 & 92 & 77 & 80 & 87 & 99 & 93 \\
Precision (\%)  & 85 & 86 & 83 & 76 & 86 & 89 & 87 & 99 & 96 \\
Specificity (\%) & 80 & 83 & 83 & 87 & 74 & 78 & 84 & 98 & 89 \\
$F_1$ Score (\%) & 85 & 86 & 84 & 84 & 80 & 84 & 87 & 99 & 95 \\
\hline\hline
\end{tabular}
\end{table*}

\section{Results}
In this work, the performance of models is obtained for cervical cancer classification. A table is used to compare easily between different models. In addition to this, the models that have the best results are highlighted in this section. All results of the metrics of models are proposed and compared in Table \ref{tab:p4}.  Firstly, in machine learning techniques; logistic regression has 83\% accuracy and recall value, 85\% precision and $F_1$ score with 80\% specificity results. For this reason, logistic regression is an average model for Herlev dataset to classify pap smear images. kNN model has 85\% accuracy, 87\% recall, 86\% precision and $F_1$ score with 83\% specificity and it is a good model. SVM has 83\% accuracy, precision and specificity with 85\% recall and 84\% $F_1$ score that is an average model. Naive Bayes classifier has 80\% accuracy and 76\% precision with the weakest results, also it has 92\% recall and 87\% specificity results with the best results. However, it is not a convenient model because of its inconsistency. Decision tree is the weakest model in machine learning algorithm with the lowest 4 metric values that are accuracy, recall, specificity and $F_1$ score. It has 80\% accuracy and $F_1$ score, 77\% recall, 86\% precision and 74\% specificity values. Random forest is an average model with 83\% accuracy, 80\% recall, 89\% precision with the highest value, 78\% senstivity and 84\% $F_1$ score. XGBoost is also the best model. It has 85\% accuracy and 87\% $F_1$ score with the highest values, 87\% recall and precision and 84\% specificity values. The best result is achieved by using XGBoost classifier with 85\% accuracy. Likewise, k-NN is the second best classifier for the pap smear images specific to Herlev dataset with 85\% accuracy. As concerns deep learning, despite it is standard CNN, its model has 99\% accuracy, recall, precision and $F_1$ score with 98\% specificity value for training data. However, a more important thing is the model performance for test data with the model. Similarly, the model has a high accuracy for test data about 93\% accuracy and recall, very good $F_1$ score with 95\% and precision with 96\% and the last metric is specificity that has 89\% value.

\section{Conclusion}
In this study, machine and deep learning methods are investigated and compared to find out which model would be an appropriate choice for cervical cell classification over Herlev dataset. A model or algorithm can be suitable for a dataset. However, if the hyperparameter of a model is not optimized for the dataset, the real performance of the model will not be seen. Thus, hyperparameter optimizations and selected features play a critical role to get a good model. The results of these models indicate that the models can obtain the highest performance with hyperparameter optimization. In addition to this, the CNN model extracts its features, and it can select more distinctive features than the hand-crafted features. The CNN model has better results than traditional machine learning techniques, as it is seen in this work.
	
\bibliographystyle{ieeetr}
\bibliography{bibliography.bib}
\end{document}